\documentstyle[preprint,aps]{revtex}
\begin{document}

\preprint{gr-qc/9307033}

\title{Dimensionally Continued Black Holes}

\author{M\'aximo Ba\~nados$^{1,2}$\thanks{Electronic address:
cecsphy@lascar.puc.cl}, Claudio Teitelboim$^{1,3,*}$
\\ and Jorge Zanelli$^{1,2,*}$}

\address{ {}~$^{1}$Centro de Estudios Cient\'{\i}ficos de Santiago, Casilla
16443, Santiago 9, Chile. \\ ~$^{2}$Departamento de F\'{\i}sica,
Facultad de Ciencias, Universidad de Chile,
\\ Casilla 653, Santiago, Chile. \\  ~$^{3}$
Institute for Advanced Study, Olden Lane, Princeton, New Jersey
08540, USA.}

\maketitle

\begin{abstract}
Static, spherically symmetric solutions of the field equations
for a particular dimensional continuation of general relativity
with negative cosmological constant are studied. The action is,
in odd dimensions, the Chern-Simons form for the anti-de Sitter
group and, in even dimensions, the Euler density constructed
with the Lorentz part of the anti-de Sitter curvature tensor.
Both actions are special cases of the Lovelock action, and they
reduce to the Hilbert action (with negative cosmological
constant) in the lower dimensional cases $\mbox{$\cal D$}=3$ and
$\mbox{$\cal D$}=4$.  Exact black hole solutions characterized
by mass ($M$) and electric charge ($Q$) are found. In odd
dimensions a negative cosmological constant is necessary to
obtain a black hole, while in even dimensions, both
asymptotically flat and asymptotically anti-de Sitter black
holes exist.  The causal structure is analyzed and the Penrose
diagrams are exhibited.  The curvature tensor is singular at the
origin for all dimensions greater than three. In dimensions of
the form $\mbox{$\cal D$}=4k,4k-1$, the number of horizons may
be zero, one or two, depending on the relative values of $M$ and
$Q$, while for negative mass there is no horizon for any real
value of $Q$.  In the other cases, $\mbox{$\cal D$}=4k+1,4k+2$,
both naked and dressed singularities with positive mass exist.
As in three dimensions, in all odd dimensions anti-de Sitter
space appears as a ``bound state" of mass $M=-1$, separated from
the continuous spectrum $(M \geq 0)$ by a gap of naked curvature
singularities.  In even dimensions anti-de Sitter space has zero
mass. The analysis is Hamiltonian throughout, considerably
simplifying the discussion of the boundary terms in the action
and the thermodynamics. The Euclidean black hole has the
topology $\Re^2 \times S^{\mbox{\tiny $\cal D$}-2}$.  Evaluation
of the Euclidean action gives explicit expressions for all the
relevant thermodynamical parameters of the system. The entropy,
defined as a surface term in the action coming from the horizon,
is shown to be a monotonically increasing function of the
black--hole radius, different from the area for $\mbox{$\cal
D$}>4$.
\end{abstract}


\section{Introduction}

The Black Hole is a central element of general relativity, but
its properties -especially at the quantum level- are far from
being completely understood. It is of interest, therefore, to
search for black holes in the extension of Einstein's
gravitation theory to higher -and lower- dimensions. Here we
consider traditional black hole solutions with
gravitational and electromagnetic fields only. Other systems,
including 2-\mbox{$\cal D$}\ black holes with a dilaton field,
and string-inspired holes, are not considered \cite{WC}.

The most general extension to higher dimensions of general
relativity, which keeps the field equations for the metric of
second order, is the so-called Lovelock action \cite{Lovelock}
which may be regarded as formed by the dimensional continuation
of the Euler characteristics of lower dimensions \cite{5}.
Although very similar in content and structure to the usual
theory, the Lovelock theory has some unwanted features.  First,
for $\mbox{$\cal D$}>4$, the time evolution of the fields
appears to be non-unique: given an initial value surface at some
time $t=t_0$, the fields at $t>t_0$ are not completely
determined by the equations of motion. This is due to the
presence of high powers of the velocities in the
Lagrangian\cite{5,7}. Second, besides Newton's constant and the
cosmological constant, the action has $n=[(D+1)/2]$ arbitrary
dimensionful parameters \cite{f1}. This arbitrariness makes the
analysis of the black--hole properties rather complicated.

In this work, we will deal neither with the pathological time
evolution problem nor with the most general set of coefficients.
Instead, we will assume from the very begining a static metric
and we will, in what we believe is a natural manner, restrict
the coefficients to a subfamily of two parameters related to
Newton's constant and to the cosmological constant. The chosen
action {\em is not} the Hilbert action with a cosmological
constant, but it reduces to it in the lower dimensional cases
$\mbox{$\cal D$}=3$ and $\mbox{$\cal D$}=4$.  We will deal only
with a negative cosmological constant related to a length
scale $l$ by $\Lambda = -a \mbox{$l^{-2}$}$ where $a$ is a
positive number. (Most of the results remain valid for positive
$\Lambda$ by the replacement $l \mbox{$\rightarrow $}il$.  In
that case, however, the spatial sections of spacetime are
closed, and no conserved quantities such as mass and electric
charge can be defined.) The construction of the action is
carried out in Sec. II.

Static spherically symmetric solutions for the Lovelock action
have been studied by several authors\cite{13,14,Myers}. However,
in those references no choice of the Lovelock coefficients is
made, making it difficult to extract physical information from
the solution. This is so because the metric components $g_{rr}$
and $g_{00}$ are the roots of a polynomial of degree $n-1$ which
cannot be solved explicitly. Appart from the
problem of solving this algebraic equation, some physical
consequences of a generic choice of the coefficients should be
remarked: (i) There exist up to $n-1$ different solutions, (ii) all of
them may have horizons for both signs of the energy, and (iii) the
entropy is not a monotonically increasing polynomial of the
horizon radius $r_+$, so that the second law of thermodynamics
does not hold. All of this suggests that a special choice of the
arbitrary coefficients should be made. The choice proposed here
produces a unique solution for the metric with a dressed
singularity only for positive masses, and the entropy is a
monotonically increasing function of $r_+$.

Our choice may not be the only one that has these properties.
But, as shown in Sec. II, it is somewhat natural in that it is
the Euler-Chern-Simons form in odd dimensions, and it may be
regarded as the gravitational analogue of the Born-Infeld
electrodynamics in even dimensions.

In Sec. III, the equations of motion are solved and explicit
expressions for the metric are obtained.  It is observed that
two branches of black holes emerge, one for even dimensions,
with strong similarities to the Schwarzschild\ metric in 3+1
dimensions, and another for odd dimensions, with many features
in common with the 2+1 black hole\cite{BTZ,BHTZ}.  The Penrose
diagrams displaying the causal structure are exhibited.

In Sec. IV the black hole thermodynamics is analyzed. It is
shown that the entropy is a surface term in the action coming
from the horizon\cite{Ent}, while the internal energy and
electric charge are given by flux integrals at infinity.  The
Hamiltonian method allows us to compute the entropy by direct
evaluation of the Euclidean action with an automatic
regularization. As it was previously noted in the
four-dimensional case\cite{York}, the Hamiltonian symplectic
structure is preserved by the thermodynamical description of the
higher dimensional black holes. That is, conjugate variables in
the sense of mechanics are also thermodynamical conjugates.

The immersion of the Lorentz Lie algebra in the anti-de Sitter
algebra is reviewed in Appendix A. The Hamiltonian structure of
Lovelock action is worked out in Ref. \cite{5} and it is
reviewed in Appendix B.


\section{Action}
\subsection{Lovelock Action}

The most general action that generalizes Einstein's gravity,
while keeping the same degrees of freedom when spacetime has a
dimension \mbox{$\cal D$}$\geq 3$, is a sum of the dimensionally
continued Euler characteristics of all dimensions below
\mbox{$\cal D$}\
\cite{Lovelock,5,f2}. The action takes the form

\begin{equation}
I= \kappa\sum_{p=0}^{n} \alpha_p I_{p}
\label{a1}
\end{equation}
with

\begin{equation}
I_p = \int \epsilon_{a_{1}...a_{D}}R^{a_{1} a_{2}} \mbox{\tiny
$\wedge$} \cdots \mbox{\tiny $\wedge$} R^{a_{2p-1}
a_{2p}}\mbox{\tiny $\wedge$} e^{a_{2p+1}} \mbox{\tiny $\wedge$}
\cdots \mbox{\tiny $\wedge$} e^{a_{D}}.
\label{a2}
\end{equation}
Here $e^a$ is the local frame 1--form, $R^{a}_{\;b} =
d\omega^{a}_{\;b} + \omega^{a}_{\;c}\mbox{\tiny
$\wedge$}\omega^{c}_{\;b}$ is the curvature 2--form, and
$\omega^{a}_{\;b}$ is the spin connection,
$a_i=\{0,1,...,\mbox{$\cal D$}-1\}$. The coefficients
$\alpha_{p}$ are arbitrary constants with dimensions
[length]$^{-(\mbox{\tiny $\cal D$}-2p)}$ and $\kappa$ has units of
action.

For even \mbox{$\cal D$}, the last term in the sum
($p=\mbox{$\cal D$}/2$) is the Euler characteristic and does not
contribute to the equations of motion.  In the quantum theory,
however, this term assigns different phases to different
topologies.

Varying (\ref{a1}) with respect to $e^a$ yields the field
equations, which are more complicated than Einstein's equations,
involving high powers of the curvature (they are non-linear in
the velocities and accelerations). These equations, however, are
still second order in the metric and the system has a standard
constrained Hamiltonian formulation\cite{5}. The variation with
respect to the spin connection $\omega$ vanishes identically by
the assumption of zero torsion.

In the standard metric formulation, the action (\ref{a1}) is
constructed by the same requirements as in $\mbox{$\cal D$}=4$
\cite{Lovelock}: general covariance, second order field
equations for the metric. In the language of forms, (\ref{a1})
can be obtained by the requirement that the Lagrangian be a
local Lorentz invariant \mbox{$\cal D$}--form (up to closed
forms), constructed entirely out of $e$, $\omega$ and their
exterior derivatives, without using the Hodge dual
(*-operation). This is equivalent, in turn, to defining the
action by dimensional continuation of topological invariants
(characteristic classes) of lower dimensions \cite{Reg,2}.

These demands, however, do not restrict the values of the
coefficients $\alpha_p$. In order to select these coefficients
we will consider embedding the Lorentz group $SO(\mbox{$\cal
D$}-1,1)$ into a larger group.  Since we are interested in
open spaces, the minimal extension for $SO(\mbox{$\cal D$}-1,1)$
is the anti-de Sitter group $SO(\mbox{$\cal D$}-1,2)$.

In odd dimensions it is possible to construct a Lagrangian
invariant under the anti-de Sitter group by making a certain
choice of the Lovelock coefficients. That Lagrangian is the
Chern-Simons form associated to the Euler density for one
dimension above $\mbox{$\cal D$}$. In even dimensions, on the
other hand, it is not possible to construct a non-trivial action
principle invariant under $SO({\cal D}-1,2)$ and it is
necessary to break the symmetry down to the Lorentz group.
This symmetry breaking is similar to that used by Mac Dowell and
Mansouri in four dimensions\cite{MS}.

In what follows, we will consider a particular choice of the
Lovelock coefficients $\alpha_p$ in (\ref{a1}) given by

\begin{equation}
\alpha_p = \left\{ \begin{array}{lll} {\displaystyle
\frac{1}{\mbox{$\cal D$}-2p}}
\left( \begin{array}{c} n-1 \\ p \end{array} \right)
l^{-\mbox{\tiny $\cal D$}
+2p} & \;\; \mbox{$\cal D$}=2n-1 & \;\;\;(a) \\ \left(
\begin{array}{c} n \\
p \end{array} \right) l^{-\mbox{\tiny $\cal D$} +2p} &
\;\;\mbox{$\cal D$}=2n & \;\;\;(b)
\end{array} \right.
\label{1.3}
\end{equation}
where $l$ is a length.  As we will see in next sections, this
choice provides an action principle with the above properties.


\subsection{Action in Odd Dimensions}

In order to construct the action for $\mbox{$\cal
D$}=2n-1$, we consider the Euler density in one dimension above
\mbox{$\cal D$}.  This density is an
exact form and can be written as

\begin{equation}
\mbox{${\cal E}$}_{2n} = \kappa \epsilon_{A_1 \cdots A_{2n}}
\tilde{R}^{A_1 A_2} \mbox{\tiny $\wedge$} \cdots
\tilde{R}^{A_{2n-1} A_{2n}} = d \mbox{\tiny $\wedge$}
\mbox{${\cal L}$}_{2n-1}
\label{1.4}
\end{equation}
For later convenience, the units are chosen so that

\begin{equation}
\kappa = \frac{l}{(\mbox{$\cal D$}-2)!\Omega_{\mbox{\tiny $\cal
D$}-2}} \;\;\;\;\;\;\;\;\;\; \mbox{($\cal D$ odd)}
\label{1.6}
\end{equation}
where $\Omega_{\mbox{\tiny $\cal D$}-2}$ is the area of the
$\mbox{$\cal D$}-2$ sphere.

The curvature tensor $\tilde{R}^{AB}$ is constructed with the
$SO(\mbox{$\cal D$}-1,2)$ connection $W^{AB}$ and the capital
latin indices run from $0$ to $2n$ (see Appendix A). The form
$\mbox{${\cal E}$}_{2n}$ cannot be used as a Lagrangian in $2n$
dimensions since it is a total derivative, but $\mbox{${\cal
L}$}_{2n-1}$ is a Lagrangian in $2n-1$ dimensions. In analogy
with the Chern-Simons forms constructed from the Pontryagin
classes, we call $\mbox{${\cal L}$}_{2n-1}$ the
Euler-Chern-Simons $(2n-1)-$form.

Using the decomposition of $W^{AB}$ into Lorentz rotations and
``inner translation" described in Appendix A which expresses the
anti-de Sitter curvature form $\tilde{R}$ in terms of the
Lorentz curvature $R$ as

\begin{equation}
\tilde{R}^{ab} = R^{ab} + \mbox{$l^{-2}$} e^a \mbox{\tiny $\wedge$} e^b,
\label{1.4-1}
\end{equation}
the following expression for $\mbox{${\cal L}$}_{2n-1}$ as a
function on the spin connection $w^{ab}$ and the local frame
$e^a$ is obtained

\begin{equation}
\mbox{${\cal L}$}_{2n-1} =  \kappa \sum_{p=0}^{n-1} \alpha_p
\epsilon_{a_{1}...a_{D}}R^{a_{1} a_{2}} \mbox{\tiny $\wedge$}
\cdots \mbox{\tiny $\wedge$}
R^{a_{2p-1} a_{2p}} \mbox{\tiny $\wedge$} e^{a_{2p+1}}
\mbox{\tiny $\wedge$} \cdots \mbox{\tiny $\wedge$}
e^{a_{\mbox{\tiny $\cal D$}}},
\label{1.5}
\end{equation}
where the $\alpha_p$ are given in (\ref{1.3}a).

Under $SO(\mbox{$\cal D$}-1,2)$ gauge transformations, the
density $\mbox{${\cal L}$}_{2n-1}$ is changed by the addition of
a closed form. This makes its integral invariant under gauge
transformations that do not change the boundary data.

The construction given here for the action in odd dimensions is
very similar to the Chern-Simons action in three dimensions. One
may ask whether the equivalence between local gauge
transformations and diffeomorphisms in three dimensions is also
valid here. The answer is in the negative because this
equivalence requires the curvature $2-$form to vanish as a
consequence of the equations of motion, which is true in three
dimensions only (see Sec. II.D).

Note that in the limit $l \rightarrow \infty$ (zero cosmological
constant), only the last term $p=n-1$ -and not the Hilbert term-
in the sum (\ref{1.5}) contributes to the action.


\subsection{ Action in Even Dimensions}

Now the situation is different: there is no analog of the
Chern-Simons action, invariant under the enlarged gauge group up
to surface terms. It is necessary to break the full anti-de
Sitter symmetry in order to produce a non-trivial action
principle \cite{MS}.

For $\mbox{$\cal D$}=2n$ the lagrangian must be of the form

\begin{equation}
\mbox{${\cal L}$}_{2n}= \kappa \tilde{R}^{A_1 A_2} \mbox{\tiny
$\wedge$} \tilde{R}^{A_3 A_4}\mbox{\tiny $\wedge$}...\mbox{\tiny
$\wedge$}
\tilde{R}^{A_{\mbox{\tiny $\cal D$}-1} A_{\mbox{\tiny $\cal
D$}}} Q_{A_1 A_2...A_{\mbox{\tiny $\cal D$}}},
\label{a5}
\end{equation}
where $Q_{A_1,A_2,...,A_{\mbox{\tiny $\cal D$}}}$ is a tensor of
rank $\mbox{$\cal D$}$ under the group and $\tilde{R}^{AB}$ is
the anti-de Sitter curvature (see Eq.  (\ref{A2}) and
(\ref{A4})).  For later convenience, we choose the units so that

\begin{equation}
\kappa = \frac{l^2}{2\mbox{$\cal D$} (\mbox{$\cal D$}-2)!
\Omega_{\mbox{\tiny $\cal D$}-2}}, \;\;\;\;\;\;\;\;\;\;\;\;\;
\mbox{($\cal D$ even)}
\label{a8}
\end{equation}
which reproduces the standard units used in \mbox{$\cal D$}=4
and provides a convenient normalization for the black hole mass
in higher dimensions.

If we choose $Q_{A_1,A_2,...,A_{\mbox{\tiny $\cal D$}}}$ as an
invariant tensor of anti-de Sitter group then, by virtue of the
Bianchi identity, the Lagrangian (\ref{a5}) gives no equations
of motion. If, instead $Q_{A_1,A_2,...,A_{\mbox{\tiny $\cal
D$}}}$ is chosen as an invariant tensor under the Lorentz group
only, namely

\begin{equation}
Q_{A_1 A_2...A_{\mbox{\tiny $\cal D$}}} = \left\{
\begin{array}{rl}
\epsilon_{a_1...a_{\mbox{\tiny $\cal D$}}} & \;\; \mbox{for} \;\;\;
A_i=a_i, \;\; (i=1,...,\mbox{$\cal D$}) \\ 0 & \;\;
\mbox{otherwise},
\end{array}  \right.
\label{a6}
\end{equation}
then (\ref{a5}) becomes

\begin{equation}
\mbox{${\cal L}$}_{2n}= \kappa (R^{a_1 a_2} + \mbox{$l^{-2}$} e^{a_1}
\mbox{\tiny $\wedge$} e^{a_2}) \mbox{\tiny $\wedge$}
\cdots\mbox{\tiny $\wedge$} (R^{a_{\mbox{\tiny $\cal D$}-1}
a_{\mbox{\tiny $\cal D$}}} + \mbox{$l^{-2}$}
e^{a_{\mbox{\tiny $\cal D$}-1}} \mbox{\tiny $\wedge$}
e^{a_{\mbox{\tiny $\cal D$}}})
\epsilon_{a_1 a_2...a_{\mbox{\tiny $\cal D$}}}
\label{a7}
\end{equation}
which corresponds to the Lagrangian in (\ref{a1}) with the
choice (\ref{1.3}b).  This lagrangian may be regarded as the
gravitational analogue of the Born-Infeld Lagrangian \cite{3}.
The analogy is more clearly brought out by rewritting (\ref{a7})
as

\begin{equation}
\mbox{${\cal L}$}_{2n}=\kappa \, Pf [R^{a b} + \mbox{$l^{-2}$} e^a
\mbox{\tiny $\wedge$} e^b]
\label{B-I}
\end{equation}
where $Pf$ denotes the Pfaffian (a multilinear function whose
square is the determinant) in the exterior product.

In the limit $l \rightarrow \infty$, the only non-vanishing
contributions to the lagrangian are the Euler density and the
term with the second highest power in the curvature tensor
($R^{n-1}$). The Euler density ($R^{n}$), which does not
contribute to the field equations, acquires in this limit an
infinite coupling constant.


\subsection{Equations of Motion}

The equations of motion derived from the Lagrangians (\ref{1.5})
and (\ref{a7}) are

\begin{equation}
(R^{a_1 a_2} + \mbox{$l^{-2}$} e^{a_1} \mbox{\tiny $\wedge$}
e^{a_2}) \mbox{\tiny $\wedge$} \cdots\mbox{\tiny $\wedge$}
(R^{a_{2n-3} a_{2n-2}} + \mbox{$l^{-2}$} e^{a_{2n-3}}
\mbox{\tiny $\wedge$} e^{a_{2n-2}}) \epsilon_{a_1
a_2...a_{2n-1}} = 0
\label{a9}
\end{equation}
in odd dimensions $(\mbox{$\cal D$}=2n-1)$, and

\begin{equation}
(R^{a_1 a_2} + \mbox{$l^{-2}$} e^{a_1} \mbox{\tiny $\wedge$}
e^{a_2}) \mbox{\tiny $\wedge$} \cdots\mbox{\tiny $\wedge$}
(R^{a_{2n-3} a_{2n-2}} + \mbox{$l^{-2}$} e^{a_{2n-3}}
\mbox{\tiny $\wedge$} e^{a_{2n-2}})\mbox{\tiny $\wedge$}
e^{a_{2n-1}}\epsilon_{a_1 a_2...a_{2n}} = 0
\label{a10}
\end{equation}
in even dimensions $(\mbox{$\cal D$}=2n)$.

As we will see later, the factorized form of these equations
-which is a consequence of the particular choice of the
coefficients- leads to a considerable simplification in the
study of the physical properties of its solutions.

Equation (\ref{a9}) represents the extension of the
Chern-Simons equation to higher dimensions.  For $\mbox{$\cal
D$} > 3$ the curvature tensor does not necessarly vanish as a
consequence of the equation of motion.  That means that the
anti-de Sitter connection is not pure gauge.  Of course
$\tilde{R}^{ab}\equiv R^{ab} + \mbox{$l^{-2}$} e^a \mbox{\tiny
$\wedge$} e^b =0 $ is a solution in both cases and it represents
anti-de Sitter space.


\section{Spherically Symmetric Solutions}


\subsection{Reduced Action and Equations of Motion}

Consider first the case of zero electric charge.  Spherically
symmetric solutions for the Einstein-Lovelock equations have
been studied by several authors\cite{13,14,Myers}.  Our goal
here is the study of those solutions in the specialized case in
which the Lovelock coefficients are chosen as in (\ref{1.3}).
For our purpose here it is enough to consider a reduced action
principle where the allowed metrics are static and spherically
symmetric. In an appropiate coordinate system the metric can be
written as

\begin{equation}
ds^2=-N^2(r)g^2(r) dt^2 + g^{-2}(r)dr^2 + r^2 d\Omega^2
\label{3.1}
\end{equation}
where $d\Omega^2 =\gamma_{m n}dx^m dx^n$ is the metric on the
$(\mbox{$\cal D$}-2)$ unit sphere (we will also use later on the
notation $\gamma = det(\gamma_{m n})$), and $N(r)$ and $g(r)$
are functions to be varied.

Working in Hamiltonian form, it is a simple matter to evaluate
the action for the metric (\ref{3.1}).  By direct application of
the results of Ref.\cite{5} one finds for the reduced action
(see Appendix B)

\begin{eqnarray}
I &=& (t_2-t_1) \int dr N \left(\frac{\mbox{$\partial $}
F}{\mbox{$\partial $} g}g' + \frac{\mbox{$\partial $}
F}{\mbox{$\partial $} r}\right) + B \nonumber \\ &=& (t_2-t_1)
\int dr N F' + B
\label{3.2}
\end{eqnarray}
Here $B$ is a surface term that will be ajusted below, the prime
denotes derivative with respect the radial coordinate $r$, and
the function $F[r,g(r)]$ is given by

\begin{equation}
F[r,g(r)]= \left\{ \begin{array}{lcr}
\frac{1}{2} r \left[ 1 + (r/l)^2 - g^2(r) \right]^{n-1} &&
\;\;\; \mbox{$\cal D$}=2n \\ \left[1 + (r/l)^2 -g^2(r)
\right]^{n-1}. && \;\;\; \mbox{$\cal D$}=2n-1
\end{array} \right.
\label{3.3}
\end{equation}
The equations of motion derived from (\ref{3.2}) are

\begin{equation}
\frac{dN}{dr}=0, \;\;\;\; \frac{dF}{dr}=0
\label{3.4}
\end{equation}
having the solutions

\begin{equation}
N(r) = N_{\infty} =const.,\;\;\;\; F[r,g(r)]=C=const.
\label{3.5}
\end{equation}

The two parameters $C$ and $\mbox{$N_{\infty}$}$ are the
constants of integration of the problem. By adjusting the time
scale, $\mbox{$N_{\infty}$}$ can be set equal to one, but it
will be conveniently keep it as an independent parameter for the
analysis of the energy and the thermodynamical properties of the
solution.  The parameter $C$ is the black-hole mass up to an
additive constant,

\begin{equation}
C=M+C_0.
\label{5.5}
\end{equation}
(see Sec. III.F). We will fix the constant $C_0$ in Sec. III.B.


\subsection{Black Holes}

{}From (\ref{3.3}) and (\ref{3.5}) we can express the metric
coefficient $g^2(r)$ as a function of the mass $M$ and $r$ as

\begin{equation}
g^2(r)=\left\{\begin{array}{llr} 1 - (2M/r)^{\frac{1}{n-1}} +
(r/l)^2 & &\; \mbox{$\cal D$}=2n \\ 1 - (M+1)^{\frac{1}{n-1}} +
(r/l)^2 & &\; \mbox{$\cal D$}=2n-1
\end{array}  \right.
\label{3.6}
\end{equation}
Choosing $\mbox{$N_{\infty}$}=1$ the metric takes the form

\begin{equation}
ds^2 = -\left[1 - (2M/r)^{\frac{1}{n-1}} + (r/l)^2\right]dt^2 +
\frac{dr^2}{1 -
(2M/r)^{\frac{1}{n-1}} + (r/l)^2} + r^2 d\Omega^2
\label{3.6a}
\end{equation}
for even dimensions, and

\begin{equation}
ds^2 = - \left[1 - (M+1)^{\frac{1}{n-1}} + (r/l)^2\right]dt^2 +
\frac{dr^2}{1 -
(M+1)^{\frac{1}{n-1}} + (r/l)^2} + r^2 d\Omega^2
\label{3.6b}
\end{equation}
for odd dimensions.

Note that for even $\cal D$ the mass has dimensions of lengh
whereas for odd $\cal D$, it is dimensionless. This is because
in the even case the constant $\kappa$ in the action has dimensions
of lengh squared (Eq. (\ref{a8})) whereas in the odd case it is
a lengh to the first power (Eq. (\ref{1.6})).

The criterion we adopt here to fix $C_0$ is that for zero energy
the horizon should disappear.  This yields

\begin{equation}
C_0 = \left\{ \begin{array}{ll} 0 & \;\;\;\;\;\;\; \mbox{$\cal D$}=2n \\
1 & \;\;\;\;\;\;\;
\mbox{$\cal D$}=2n-1. \end{array}
\right.
\label{3.7-1}
\end{equation}

The usual cases \mbox{$\cal D$}=4 and \mbox{$\cal D$}=3 are
obtained from (\ref{3.6a}) and (\ref{3.6b}) in the special case
$n=2$.

The metrics (\ref{3.6a}) and (\ref{3.6b}) describe a black hole
if the function $g^2(r)$ has at least one root for a real
positive value $r_+$,

\begin{equation}
g^2(r_+)=0, \;\;\; r_+>0.
\label{3.7}
\end{equation}

{}From (\ref{3.6}) it follows that for odd $n$ ($\mbox{$\cal
D$}=4k+1,4k+2$), the following happens: (i) there is a sign
ambiguity in the $(n-1)$-th root that appears in $g^2$ and (ii)
the mass $M$ must be positive in order to have a real solution.
If the $+$ sign is chosen, then the solution has an event
horizon. If the minus sign is chosen, the solution has a naked
singularity.  Note that one could have a naked singularity with
positive mass in this case.

For even $n$ $(\mbox{$\cal D$}=4k-1,4k)$, there is no sign
ambiguity and it is a simple matter to check that a horizon
exists if and only if the mass is positive.

We will take as a fundamental requirement that there should be
no naked singularities with positive mass. This is a form of
cosmic censorship. Hence we will exclude odd $n$ from the
physical spectrum and we are only left with the following spacetime
dimensions

$$\begin{array}{lcccr}
\mbox{$\cal D$} &=& 4k-1 &=& 3,7,11,...\\
\mbox{$\cal D$} &=& 4k   &=& 4,8,12,... \end{array} $$


\subsection{Causal Structure in Even Dimensions}

The metric (\ref{3.6a}) for $\mbox{$\cal D$}=4k$ represents
a natural extension of the Schwarzschild geometry to higher
dimensions.  The metric diverges at the origin as
$r^{\frac{1}{2k-1}}$ and the curvature tensor is also singular
there.  The asymptotic region, on the other hand, approaches
anti-de Sitter space.  The Penrose diagram of that solution does
not depend on \mbox{$\cal D$}, and it is shown in Fig. 1a.
There are two asymptotically anti-de Sitter regions (vertical
lines) and two singularities, past and future, (horizontal
lines) where the geodesics end. The horizon is drawn as $45^o$
lines.


\setlength{\unitlength}{1mm}

\begin{center}
\begin{picture}(100,50)
\footnotesize
\put(10,40){\linethickness{.7mm}\line(1,0){30}}
\put(10,10){\linethickness{.7mm}\line(1,0){30}}
\put(10,10){\line(0,1){30}}
\put(40,10){\line(0,1){30}}
\put(10,10){\line(1,1){30}}
\put(10,40){\line(1,-1){30}}
\put(25,8){\makebox(0,0){$r=0$}}
\put(25,42){\makebox(0,0){$r=0$}}
\put(5,25){\makebox(0,0){$r=\infty$}}
\put(46,25){\makebox(0,0){$r=\infty$}}
\put(34,34){\makebox(0,0){$r=r_+$}}
\put(34,16){\makebox(0,0){$r=r_+$}}
\put(31,25){\makebox(0,0){\mbox{${\bf I}$}}}
\put(25,31){\makebox(0,0){\mbox{${\bf I}$}\mbox{${\bf I}$}}}
\put(19,25){\makebox(0,0){\mbox{${\bf I}$}\mbox{${\bf I}$}\mbox{${\bf I}$}}}
\put(25,19){\makebox(0,0){\mbox{${\bf I}$}{\bf V}}}
\put(66,10){\linethickness{.7mm}\line(0,1){30}}
\put(91,10){\line(0,1){30}}
\put(61,25){\makebox(0,0){$r=0$}}
\put(98,25){\makebox(0,0){$r=\infty$}}
\put(77,7){\makebox(0,0){$i^- \bullet$}}
\put(77,43){\makebox(0,0){$i^+ \bullet$}}
\end{picture}\\

(a) $M > 0$ \hspace{3.2cm} (b) $M=0$

\noindent Fig. 1 \\
Penrose Diagram for \mbox{$\cal D$}=2n \\
\end{center}

In the case of zero mass one obtains anti-de Sitter space (Fig.
1b).  The points $i^+$ and $i^-$ represents future and past
infinity and are disjoint points in the figure. Further
properties of that space can be found in Ref.
\cite{Hawking-Ellis}.


\subsection{Causal Structure in Odd Dimensions}

The higher dimensional black hole in odd dimensions has a strong
resemblance with the 2+1 black hole solution\cite{BTZ}.
However, for $\mbox{$\cal D$}>3$ the curvature tensor is no
longer constant as in $\mbox{$\cal D$}=3$ and, therefore, those
solutions cannot be obtained from anti-de Sitter space by an
identification process\cite{BHTZ}. The non vanishing components
of the curvature tensor are

\begin{equation}
\begin{array}{rcl}
R^{0r}_{0r} &=& -\mbox{$l^{-2}$} \\ R^{0 m}_{0 n} &=&
-\mbox{$l^{-2}$} \delta^m_n \\ R^{r m}_{r n} &=&
-\mbox{$l^{-2}$} \delta^m_n \\ R^{m_1 m_2}_{n_1 n_2} &=& \left[
-\mbox{$l^{-2}$} + \frac{(M+1)^{\frac{1}{n-1}}}{r^2}\right]
\delta^{[m_1
m_2]}_{[n_1 n_2]}
\end{array}
\label{3.8}
\end{equation}
where the antisymmetrized delta function is normalized by
$\delta^{[m_1 m_2]}_{[m_1 m_2]}= (\mbox{$\cal D$}-2)(\mbox{$\cal
D$}-3)$.

The scalar curvature is

\begin{equation}
R=-\frac{\mbox{$\cal D$} (\mbox{$\cal D$}-1)}{l^2} +
\frac{(M+1)^{\frac{1}{n-1}}
(\mbox{$\cal D$}-2)(\mbox{$\cal D$}-3)}{r^2}. \;\;\;\;\;\;\;
(\mbox{$\cal D$}=2n-1)
\label{3.9}
\end{equation}

As we showed before, for negative values of $M$ there are no
horizons and a naked singularity is obtained.  However, when one
reaches the point $M=-1$ the singularity in the curvature tensor
vanishes and the geometry is regular; the space obtained has
constant negative curvature (anti-de Sitter space).  This is
quite similar to the lower dimensional case of the black hole in
three dimensions. In that case, a naked conical singularity develops for
negative masses, but when one reaches $M=-1$ the singularity
disappears and the solution also becomes anti-de Sitter
space.  In this sense anti-de Sitter space emerges as a ``bound
state" in the mass spectrum\cite{BTZ}.

At large distances from $r=0$, $R^{\alpha \beta}_{\;\;
\mu \nu} \mbox{$\rightarrow $}-l^{-2} \delta^{[\alpha
\beta]}_{[\mu \nu]}$ and the spacetime approaches anti de Sitter space.

Note also that for $\mbox{$\cal D$}=3$ the curvature singularity
disappears for any value of $M$, and one obtains a space of
constant negative curvature.  As it was shown in \cite{BHTZ} the
black hole spacetime may then be obtained from anti-de Sitter
space through identifications.  The surface $r=0$ is then a
singularity in the causal structure only, in that closed timelike
lines appear for $r<0$. It was argued in
\cite{BHTZ} that the metric smoothness at $r=0$ is unstable
under matter couplings. The present results show that it is also
unstable upon dimensional continuation.

The Penrose diagram in odd dimensions is shown in figure 2.
Three different types of states exist in this case, (a) the
generic case $M \geq 0$, (b) the vacuum state $M=0$ and (c)
anti-de Sitter space of mass $M=-1$.


\setlength{\unitlength}{1mm}

\begin{center}
\begin{picture}(140,50)
\footnotesize
\put(10,40){\linethickness{.7mm}\line(1,0){30}}
\put(10,10){\linethickness{.7mm}\line(1,0){30}}
\put(10,10){\line(0,1){30}}
\put(40,10){\line(0,1){30}}
\put(10,10){\line(1,1){30}}
\put(10,40){\line(1,-1){30}}
\put(25,8){\makebox(0,0){$r=0$}}
\put(25,42){\makebox(0,0){$r=0$}}
\put(5,25){\makebox(0,0){$r=\infty$}}
\put(46,25){\makebox(0,0){$r=\infty$}}
\put(34,34){\makebox(0,0){$r=r_+$}}
\put(34,16){\makebox(0,0){$r=r_+$}}
\put(31,25){\makebox(0,0){\mbox{${\bf I}$}}}
\put(25,31){\makebox(0,0){\mbox{${\bf I}$}\mbox{${\bf I}$}}}
\put(19,25){\makebox(0,0){\mbox{${\bf I}$}\mbox{${\bf I}$}\mbox{${\bf I}$}}}
\put(25,19){\makebox(0,0){\mbox{${\bf I}$}{\bf V}}}
\put(65,10){\line(0,1){30}}
\put(65,10){\line(1,1){15}}
\put(65,40){\line(1,-1){15}}
\put(59,25){\makebox(0,0){$r=\infty$}}
\put(76,15){\makebox(0,0){$r=0$}}
\put(76,35){\makebox(0,0){$r=0$}}
\put(100,10){\linethickness{.7mm}\line(0,1){30}}
\put(125,10){\line(0,1){30}}
\put(95,25){\makebox(0,0){$r=0$}}
\put(132,25){\makebox(0,0){$r=\infty$}}
\put(111,7){\makebox(0,0){$i^- \bullet$}}
\put(111,43){\makebox(0,0){$i^+ \bullet$}}
\end{picture}\\

(a) $M > 0$ \hspace{2.8cm} (b) $M=0$ \hspace{2.4cm} (c) $M=-1$

\noindent Fig. 2 \\
Penrose Diagrams in $\mbox{$\cal D$}=2n-1$ \\
\end{center}


\subsection{Charged Black Holes}

Electric charge can be incorporated by coupling the
electromagnetic field to the gravitational part of the action.

The Hamiltonian action for the electromagnetic field in a curved
background is

\begin{equation}
I_{elm} = \int dt \int d^{\mbox{\tiny ${\cal D}$}-1}x \left[ p^i
\dot{A}_i - \frac{1}{2} \mbox{$N^{\perp}$} \left(\alpha g^{-1/2}
p^i p_i + \frac{g^{1/2}}{2\alpha} F^{ij}F_{ij}\right) + \varphi
p^i,_i \right] + B_{elm}
\label{q1}
\end{equation}
where $P^i$ is the momentum conjugate to the spatial components
of the gauge field $A_i$, $\varphi \equiv A_0$, $B_{elm}$ is a
surface term that depends on the boundary conditions, and $\alpha $
may be conveniently taken to be equal to the area of the
$(\mbox{$\cal D$}-2)$ unit sphere. We will only be interested in
solutions without magnetic charge, static and spherically
symmetric, that is,

\begin{equation}
\begin{array}{rclcll}
F_{ij} &=& 0 & & &\;\;\; \mbox{(no magnetic field)} \\ p^i &=&
(0,p^r,0,...,0) & & &\;\;\; \mbox{(radial field)} \\
\dot{A}_i &=& 0 = \dot{p}^i & &   &\;\;\; \mbox{(static field)}
\end{array}
\label{q2}
\end{equation}

We may impose the conditions (\ref{q2}) and then proceed to
vary the action. The reduced action for the Coulomb field takes
de form

\begin{equation}
I_{elm}^{red.} = (t_2-t_1)\int dr \left[ -\frac{1}{2} N
r^{\mbox{\tiny $\cal D$}-2} p^2 +
\varphi (r^{\mbox{\tiny $\cal D$}-2}p)'\right] + B_{elm}.
\label{q3}
\end{equation}
where $N=\mbox{$N^{\perp}$} g_{rr}^{1/2}$ and $p$ is the
rescaled radial component of $p^i$

\begin{equation}
p^r = r^{\mbox{\tiny $\cal D$}-2} \frac{\gamma^{1/2}}{\alpha} p
\label{q4}
\end{equation}

The total reduced action of the gravitational plus
electromagnetic system is then

\begin{eqnarray}
I &=& I_G + I_{elm} \nonumber \\ &=& (t_2-t_1) \int dr \left[ N
(F'- \frac{1}{2} r^{\mbox{\tiny $\cal D$}-2} p^2 ) + \varphi
(r^{\mbox{\tiny $\cal D$}-2}p)' \right] + B.
\label{q5}
\end{eqnarray}
where the function $F[r,g(r)]$ is defined by (\ref{3.3}) and $B$
is a boundary term that will be fixed in next section.

Varying the action (\ref{q5}) with respect to $N,g,P$ and
$\varphi$ the following equations are found

\begin{eqnarray}
F' &=& \frac{1}{2} r^{\mbox{\tiny $\cal D$}-2} p^2 \label{q6} \\
(r^{\mbox{\tiny $\cal D$}-2}p)' &=& 0
\label{q7} \\
\varphi' &=& -N p \label{q8} \\ N' &=& 0 \label{q9}
\end{eqnarray}
whose solutions are

\begin{eqnarray}
p &=& \frac{Q}{r^{\mbox{\tiny $\cal D$}-2}} \label{q10} \\
\varphi &=& \frac{\mbox{$N_{\infty}$} Q}{(\mbox{$\cal D$}-3)
r^{\mbox{\tiny $\cal D$}-3}}+\varphi_{\infty}\label{q11}\\
F &=& -\frac{\frac{1}{2} Q^2}{(\mbox{$\cal D$}-3)r^{\mbox{\tiny
$\cal D$}-3}} + C \label{q12} \\ N &=&
\mbox{$N_{\infty}$} \label{q13}.
\end{eqnarray}
The parameters $Q,C=M+C_0,\varphi_{\infty}$ and
\mbox{$N_{\infty}$}\ are the
integration constants of the problem; $\varphi_{\infty}$ and
\mbox{$N_{\infty}$}\
are the values of $\varphi$ and $N$ at infinity and are
conjugates to $Q$ and $M$ respectively (see Sec. III.F).  From
(\ref{q12}) and (\ref{3.3}), choosing \mbox{$N_{\infty}$}=1, one
can obtain the form of the metric,

\begin{equation}
ds^2=-g^2(r) dt^2 + g^{-2}(r)dr^2 + r^2 d\Omega^2
\label{q14}
\end{equation}
with
\begin{equation}
g^2(r)=\left\{ \begin{array}{lll} 1 + \frac{r^2}{l^2} - \left[
\frac{2M}{r} - \frac{Q^2}{(\mbox{\tiny $\cal D$}-3)
r^{\mbox{\tiny $\cal D$}-2}} \right]^{ \frac{1}{n-1}} &\;\;\;\;
(\mbox{$\cal D$}=2n) & \\ 1 +
\frac{r^2}{l^2} - \left[ M+1 - \frac{Q^2}{2(\mbox{\tiny $\cal
D$}-3) r^{\mbox{\tiny $\cal D$}-3}}\right]^{\frac{1}{n-1}}
&\;\;\;\; (\mbox{$\cal D$}=2n-1) & \end{array}  \right.
\label{q15}
\end{equation}

For each of the two functions given by (\ref{q15}), the equation
$g^2=0$ has two, one or zero solutions depending on whether for
a given mass $M$ the squared charge is greater, equal or less
than the extremal value $Q^2_{ext}$. It is not possible to obtain an
expression for $Q^2_{ext}$ in closed form but it may be shown to
exist by a graphical analysis. Thus, charged black holes only
exist for $Q^2 < Q^2_{ext}$. Of the two roots $r_+$ and $r_-$
that exist when $Q^2 < Q^2_{ext}$, the greater root $r_+$ is the
black-hole horizon. Both roots coalesce when $Q^2=Q^2_{ext}$
corresponding to the extremal black hole.

The charged black-hole geometry has curvature singularities: $r=0$, and
$r=r_c$, where the expression whose $(n-1)$-th root
appears in (\ref{q15}) vanishes. This may be seen by looking at
the curvature scalar of the metric (\ref{q14}) which reads

\begin{equation}
\tilde{R} = \frac{1}{r^{\mbox{\tiny $\cal D$}-2}} \left[
r^{\mbox{\tiny $\cal D$}-2}(1-g^2) \right]^{''}.
\label{q16}
\end{equation}

The curvature singularity $r=r_c$, which only exists in the
charged case, is also hidden by the horizon. Indeed one has

$$ 0 < r_c < r_- < r_+ $$
as may be again verified by a graphical analysis of $g^2$.

In the generic case, when two horizons are present, the manifold
splits into three differents regions: region \mbox{${\bf I}$}\ or
outer region ($r>r_+$), the intermediate region \mbox{${\bf
I}$}\mbox{${\bf I}$}\ ($r_-<r<r_+$), and the inner region
\mbox{${\bf I}$}\mbox{${\bf I}$}\mbox{${\bf I}$}\ ($r<r_-$). The
causal connection between these regions is shown in the Penrose
diagram in Fig 3.

\begin{center}
\setlength{\unitlength}{1.1mm}
\begin{picture}(140,50)
\put(10,10){\line(1,0){120}}
\put(10,40){\line(1,0){120}}
\put(10,10){\linethickness{.7mm}\line(1,0){30}}
\put(70,10){\linethickness{.7mm}\line(1,0){30}}
\put(10,40){\linethickness{.7mm}\line(1,0){30}}
\put(70,40){\linethickness{.7mm}\line(1,0){30}}
\put(10,10){\line(1,1){30}}
\put(40,10){\line(1,1){30}}
\put(70,10){\line(1,1){30}}
\put(100,10){\line(1,1){30}}
\put(10,40){\line(1,-1){30}}
\put(40,40){\line(1,-1){30}}
\put(70,40){\line(1,-1){30}}
\put(100,40){\line(1,-1){30}}
\footnotesize
\put(25,7){\makebox(0,0){$r=0$}}
\put(55,7){\makebox(0,0){$r=\infty$}}
\put(85,7){\makebox(0,0){$r=0$}}
\put(115,7){\makebox(0,0){$r=\infty$}}
\put(25,43){\makebox(0,0){$r=0$}}
\put(55,43){\makebox(0,0){$r=\infty$}}
\put(85,43){\makebox(0,0){$r=0$}}
\put(115,43){\makebox(0,0){$r=\infty$}}
\put(36,34){\makebox(0,0){$r=r_-$}}
\put(36,16){\makebox(0,0){$r=r_-$}}
\put(64,34){\makebox(0,0){$r=r_+$}}
\put(64,16){\makebox(0,0){$r=r_+$}}
\put(96,34){\makebox(0,0){$r=r_-$}}
\put(96,16){\makebox(0,0){$r=r_-$}}
\put(124,34){\makebox(0,0){$r=r_+$}}
\put(124,16){\makebox(0,0){$r=r_+$}}
\put(25,37){\makebox(0,0){$r=r_c$}}
\put(25,12){\makebox(0,0){$r=r_c$}}
\put(85,37){\makebox(0,0){$r=r_c$}}
\put(85,12){\makebox(0,0){$r=r_c$}}
\put(10,10){\line(3,1){6}} \put(16,12){\line(6,1){6}}
\put(22,13){\line(6,0){6}} \put(28,13){\line(6,-1){6}}
\put(34,12){\line(3,-1){6}}
\put(70,10){\line(3,1){6}} \put(76,12){\line(6,1){6}}
\put(82,13){\line(6,0){6}} \put(88,13){\line(6,-1){6}}
\put(94,12){\line(3,-1){6}}
\put(10,40){\line(3,-1){6}} \put(16,38){\line(6,-1){6}}
\put(22,37){\line(6,0){6}} \put(28,37){\line(6,1){6}}
\put(34,38){\line(3,1){6}}
\put(70,40){\line(3,-1){6}} \put(76,38){\line(6,-1){6}}
\put(82,37){\line(6,0){6}} \put(88,37){\line(6,1){6}}
\put(94,38){\line(3,1){6}}
\put(15,25){\makebox(0,0){\mbox{${\bf I}$}\mbox{${\bf I}$}}}
\put(35,25){\makebox(0,0){\mbox{${\bf I}$}\mbox{${\bf I}$}}}
\put(75,25){\makebox(0,0){\mbox{${\bf I}$}\mbox{${\bf I}$}}}
\put(95,25){\makebox(0,0){\mbox{${\bf I}$}\mbox{${\bf I}$}}}
\put(125,25){\makebox(0,0){\mbox{${\bf I}$}\mbox{${\bf I}$}}}
\put(25,31){\makebox(0,0){\mbox{${\bf I}$}\mbox{${\bf I}$}\mbox{${\bf I}$}}}
\put(25,19){\makebox(0,0){\mbox{${\bf I}$}\mbox{${\bf I}$}\mbox{${\bf I}$}}}
\put(55,31){\makebox(0,0){\mbox{${\bf I}$}}}
\put(55,19){\makebox(0,0){\mbox{${\bf I}$}}}
\put(85,31){\makebox(0,0){\mbox{${\bf I}$}\mbox{${\bf I}$}\mbox{${\bf I}$}}}
\put(85,19){\makebox(0,0){\mbox{${\bf I}$}\mbox{${\bf I}$}\mbox{${\bf I}$}}}
\put(115,31){\makebox(0,0){\mbox{${\bf I}$}}}
\put(115,19){\makebox(0,0){\mbox{${\bf I}$}}}
\end{picture}

\noindent Fig. 3

Penrose diagram for the charged black hole \\ (generic case)

\end{center}

For the extreme case (one horizon), there exist only two
regions, inner (\mbox{${\bf I}$}), $r<r_+$ and outer
(\mbox{${\bf I}$}\mbox{${\bf I}$}), $r>r_+$; the Penrose diagram
is shown in Fig. 4

\setlength{\unitlength}{1mm}
\begin{center}
\begin{picture}(140,50)
\put(10,10){\linethickness{.7mm}\line(1,0){120}}
\put(10,40){\line(1,0){120}}
\put(10,10){\line(1,1){30}}
\put(70,10){\line(1,1){30}}
\put(40,40){\line(1,-1){30}}
\put(100,40){\line(1,-1){30}}
\put(10,10){\line(3,1){12}}
\put(22,14.1){\line(6,1){12}}
\put(34,16.1){\line(6,0){12}}
\put(46,16.1){\line(6,-1){12}}
\put(58,14.1){\line(3,-1){12}}
\put(70,10){\line(3,1){12}}
\put(82,14.1){\line(6,1){12}}
\put(94,16.1){\line(6,0){12}}
\put(106,16.1){\line(6,-1){12}}
\put(118,14.1){\line(3,-1){12}}
\put(70,7){\makebox(0,0){$r=0$}}
\put(70,43){\makebox(0,0){$r=\infty$}}
\put(23,20){\makebox(0,0){$r=r_+$}}
\put(58,20){\makebox(0,0){$r=r_+$}}
\put(83,20){\makebox(0,0){$r=r_+$}}
\put(119,20){\makebox(0,0){$r=r_+$}}
\put(40,14){\makebox(0,0){$r=r_c$}}
\put(100,14){\makebox(0,0){$r=r_c$}}
\put(40,28){\makebox(0,0){\mbox{${\bf I}$}\mbox{${\bf I}$}}}
\put(70,28){\makebox(0,0){\mbox{${\bf I}$}}}
\put(100,28){\makebox(0,0){\mbox{${\bf I}$}\mbox{${\bf I}$}}}
\end{picture}

\noindent Fig. 4

Penrose diagram for the charged black hole \\ (extreme case)

\end{center}


\subsection{Surface Integrals: Mass and Electric Charge}

The black-hole mass and electric charge can be identified in a
simple manner by using the Hamiltonian approach.  The
surface term $B$ present in (\ref{q5}) must be chosen so that
the action has an extremum under variations of the fields with
appropiate boundary conditions. One demands that the fields
approach the classical solutions (equations
(\ref{q10})-(\ref{q13})) at infinity \cite{11}. Varying the
action (\ref{q5}), one obtains for the boundary term $B$,

\begin{equation}
\delta B = (t_2-t_1) (-\mbox{$N_{\infty}$} \delta M -
\varphi_{\infty} \delta Q ).
\label{m1}
\end{equation}

The term $B$ is the conserved charge associated to the
``improper gauge tranformations" produced by time evolution
\cite{BCT}. These transformations are: time displacements, whose
charge is the mass $(M)$, and asympotically constant gauge
transformations of the electromagnetic field, whose conserved charge is
the electric charge $(Q)$. From equation (\ref{m1}) one learns
that $(\mbox{$N_{\infty}$},M)$ and $(\varphi_{\infty},Q)$ are
conjugate pairs.  Therefore, if $M$ and $Q$ are varied, their
conjugates, $\mbox{$N_{\infty}$}$ and $\varphi_{\infty}$, must
be fixed. Thus, the boundary term is

\begin{equation}
B = (t_2-t_1) (-\mbox{$N_{\infty}$} M - \varphi_{\infty} Q ) +
B_0
\label{m2}
\end{equation}
where $B_0$ is an arbitrary constant without variation. The
freedom in the value of this constant was used in Secs. III.B
and III.E to fix the zero-point energy when the horizon
disappears.

In asympotically anti-de Sitter spaces, the rescaled lapse
function $N \equiv \mbox{$N^{\perp}$} g^{-1}$, measures time
displacements at infinity along the Killing vector
$\mbox{$\partial $}/\mbox{$\partial $} t$, instead of the usual
lapse function $\mbox{$N^{\perp}$}$. This is so because the
normal vector {\bf n} = $\mbox{$N^{\perp}$}\mbox{$\partial
$}/\mbox{$\partial $} t$ does not approach the Killing vector
$\mbox{$\partial $}/\mbox{$\partial $} t$ at infinity since
\mbox{$N^{\perp}$}\ diverges there. The rescaled vector
$\tilde{{\bf n}}$=$g^{-1} \mbox{$N^{\perp}$} \mbox{$\partial
$}/\mbox{$\partial $} t=N \mbox{$\partial $}/\mbox{$\partial $}
t$ does approach the Killing vector up to a constant factor.


\subsection{Asymptotically Flat Black Hole in Even Dimensions}

So far we have only dealt with asymptotically anti-de Sitter
spaces. In odd dimensions one is forced to deal with this
asymptotic form since no black-hole horizons are found in the
absence of cosmological constant, even in the presence of an
electric charge \cite{f3}.

In even dimensions, on the other hand, one can take the limit $l
\rightarrow \infty$ in the solution (\ref{q15}) obtaining the asymptotically
flat line element

\begin{equation}
ds^2=- \left\{1 - \left[ \frac{2M}{r} - \frac{Q^2}{(\mbox{$\cal
D$}-3) r^{\mbox{\tiny $\cal D$}-2}} \right]^{
\frac{1}{n-1}}\right\} dt^2 + \frac{dr^2}{1
- \left[ \frac{2M}{r} - \frac{Q^2}{({\cal D}-3)
r^{\mbox{\tiny $\cal D$}-2}} \right]^{
\frac{1}{n-1}}} + r^2 d\Omega^2.
\label{f1}
\end{equation}
In this limit, $M$ becomes the usual mass conjugate to
asymptotic displacements in proper time.

The black-hole radius is related to the mass and electric charge
by the equation

\begin{equation}
r_+^{\mbox{\tiny $\cal D$}-2} - 2M r_+^{\mbox{\tiny $\cal D$}-3}
+ \frac{Q^2}{\mbox{$\cal D$}-3} = 0,
\label{f3}
\end{equation}
which has zero, one or two solutions, depending on the relative
values of the mass and charge.  The Penrose diagram for the
generic case (two horizons) is shown in Fig. 5.

\begin{center}
\setlength{\unitlength}{1.1mm}
\begin{picture}(140,80)
\put(10,25){\linethickness{.7mm}\line(1,0){30}}
\put(70,25){\linethickness{.7mm}\line(1,0){30}}
\put(10,55){\linethickness{.7mm}\line(1,0){30}}
\put(70,55){\linethickness{.7mm}\line(1,0){30}}
\put(130,25){\linethickness{.7mm}\line(1,0){5}}
\put(130,55){\linethickness{.7mm}\line(1,0){5}}
\put(10,25){\line(1,1){45}}
\put(40,25){\line(1,1){30}}
\put(55,10){\line(1,1){60}}
\put(100,25){\line(1,1){30}}
\put(115,10){\line(1,1){15}}
\put(10,55){\line(1,-1){45}}
\put(40,55){\line(1,-1){30}}
\put(55,70){\line(1,-1){60}}
\put(100,55){\line(1,-1){30}}
\put(115,70){\line(1,-1){15}}
\footnotesize
\put(25,22){\makebox(0,0){$r=0$}}
\put(85,22){\makebox(0,0){$r=0$}}
\put(47,15){\makebox(0,0){${\it I}^+$}}
\put(64,15){\makebox(0,0){${\it I}^-$}}
\put(107,15){\makebox(0,0){${\it I}^+$}}
\put(124,15){\makebox(0,0){${\it I}^-$}}
\put(25,58){\makebox(0,0){$r=0$}}
\put(85,58){\makebox(0,0){$r=0$}}
\put(46,63){\makebox(0,0){${\it I}^+$}}
\put(66,63){\makebox(0,0){${\it I}^-$}}
\put(106,63){\makebox(0,0){${\it I}^+$}}
\put(126,63){\makebox(0,0){${\it I}^-$}}
\put(36,49){\makebox(0,0){$r=r_-$}}
\put(36,31){\makebox(0,0){$r=r_-$}}
\put(64,49){\makebox(0,0){$r=r_+$}}
\put(64,31){\makebox(0,0){$r=r_+$}}
\put(96,49){\makebox(0,0){$r=r_-$}}
\put(96,31){\makebox(0,0){$r=r_-$}}
\put(124,49){\makebox(0,0){$r=r_+$}}
\put(124,31){\makebox(0,0){$r=r_+$}}
\put(25,52){\makebox(0,0){$r=r_c$}}
\put(25,27){\makebox(0,0){$r=r_c$}}
\put(85,52){\makebox(0,0){$r=r_c$}}
\put(85,27){\makebox(0,0){$r=r_c$}}
\put(10,25){\line(3,1){6}} \put(16,27){\line(6,1){6}}
\put(22,28){\line(6,0){6}} \put(28,28){\line(6,-1){6}}
\put(34,27){\line(3,-1){6}}
\put(70,25){\line(3,1){6}} \put(76,27){\line(6,1){6}}
\put(82,28){\line(6,0){6}} \put(88,28){\line(6,-1){6}}
\put(94,27){\line(3,-1){6}}
\put(10,55){\line(3,-1){6}} \put(16,53){\line(6,-1){6}}
\put(22,52){\line(6,0){6}} \put(28,52){\line(6,1){6}}
\put(34,53){\line(3,1){6}}
\put(70,55){\line(3,-1){6}} \put(76,53){\line(6,-1){6}}
\put(82,52){\line(6,0){6}} \put(88,52){\line(6,1){6}}
\put(94,53){\line(3,1){6}}
\put(15,40){\makebox(0,0){\mbox{${\bf I}$}\mbox{${\bf I}$}}}
\put(35,40){\makebox(0,0){\mbox{${\bf I}$}\mbox{${\bf I}$}}}
\put(75,40){\makebox(0,0){\mbox{${\bf I}$}\mbox{${\bf I}$}}}
\put(95,40){\makebox(0,0){\mbox{${\bf I}$}\mbox{${\bf I}$}}}
\put(125,40){\makebox(0,0){\mbox{${\bf I}$}\mbox{${\bf I}$}}}
\put(25,46){\makebox(0,0){\mbox{${\bf I}$}\mbox{${\bf I}$}\mbox{${\bf I}$}}}
\put(25,34){\makebox(0,0){\mbox{${\bf I}$}\mbox{${\bf I}$}\mbox{${\bf I}$}}}
\put(55,55){\makebox(0,0){\mbox{${\bf I}$}}}
\put(55,25){\makebox(0,0){\mbox{${\bf I}$}}}
\put(85,46){\makebox(0,0){\mbox{${\bf I}$}\mbox{${\bf I}$}\mbox{${\bf I}$}}}
\put(85,34){\makebox(0,0){\mbox{${\bf I}$}\mbox{${\bf I}$}\mbox{${\bf I}$}}}
\put(115,55){\makebox(0,0){\mbox{${\bf I}$}}}
\put(115,25){\makebox(0,0){\mbox{${\bf I}$}}}
\end{picture}

\noindent Fig. 5

Penrose diagram of the asymptotically flat \\ charged black hole
in even dimensions \\ (non-extreme case) \\

\end{center}

The extreme case (one horizon) ocurs for the value

\begin{equation}
Q^2_{ext} = a_n M^n
\label{f4}
\end{equation}
where

$$ a_n = (2n-3)(n-1)^{n-1} \left(\frac{2}{n}\right)^n $$

For a given $M$ two horizons exist if $Q^2 < Q^2_{ext}$. They
coalesce if $Q^2 = Q^2_{ext}$ and if $Q^2 > Q^2_{ext}$ there is no
horizon.

The Penrose diagram for the extreme case is shown in Fig. 6.


\begin{center}

\setlength{\unitlength}{.8mm}
\begin{picture}(140,80)
\put(10,10){\linethickness{.7mm}\line(1,0){120}}
\put(10,10){\line(1,1){60}}
\put(70,10){\line(1,1){60}}
\put(10,70){\line(1,-1){60}}
\put(70,70){\line(1,-1){60}}
\put(10,10){\line(3,1){12}}
\put(22,14.1){\line(6,1){12}}
\put(34,16.1){\line(6,0){12}}
\put(46,16.1){\line(6,-1){12}}
\put(58,14.1){\line(3,-1){12}}
\put(70,10){\line(3,1){12}}
\put(82,14.1){\line(6,1){12}}
\put(94,16.1){\line(6,0){12}}
\put(106,16.1){\line(6,-1){12}}
\put(118,14.1){\line(3,-1){12}}
\put(70,7){\makebox(0,0){$r=0$}}
\put(29,57){\makebox(0,0){${\it I}^-$}}
\put(54,57){\makebox(0,0){${\it I}^+$}}
\put(89,57){\makebox(0,0){${\it I}^-$}}
\put(115,57){\makebox(0,0){${\it I}^+$}}
\put(23,20){\makebox(0,0){$r=r_+$}}
\put(58,20){\makebox(0,0){$r=r_+$}}
\put(83,20){\makebox(0,0){$r=r_+$}}
\put(119,20){\makebox(0,0){$r=r_+$}}
\put(40,14){\makebox(0,0){$r=r_c$}}
\put(100,14){\makebox(0,0){$r=r_c$}}
\put(40,28){\makebox(0,0){\mbox{${\bf I}$}\mbox{${\bf I}$}}}
\put(70,40){\makebox(0,0){\mbox{${\bf I}$}}}
\put(15,40){\makebox(0,0){\mbox{${\bf I}$}}}
\put(125,40){\makebox(0,0){\mbox{${\bf I}$}}}
\put(100,28){\makebox(0,0){\mbox{${\bf I}$}\mbox{${\bf I}$}}}
\end{picture}

\noindent Fig. 6

Penrose diagram for the asymptotically flat \\ charged black
hole in even dimensions \\ (extreme case) \\

\end{center}

In the case of zero electric charge one obtains the
Schwarzschild-like line element

\begin{equation}
ds^2=- \left\{1 - \left[ \frac{2M}{r}
\right]^{\frac{1}{n-1}}\right\} dt^2 + \frac{dr^2}{1 - \left[
\frac{2M}{r} \right]^{ \frac{1}{n-1}}} + r^2 d\Omega^2
\label{f5}
\end{equation}
which has only one horizon at $r_+=2M$. The Penrose
diagram for this case is shown in Fig. 7.

\begin{center}
\setlength{\unitlength}{1.1mm}

\begin{picture}(140,50)(20,0)
\put(70,10){\linethickness{.7mm}\line(1,0){30}}
\put(70,40){\linethickness{.7mm}\line(1,0){30}}
\put(55,25){\line(1,1){15}}
\put(55,25){\line(1,-1){15}}
\put(100,10){\line(1,1){15}}
\put(100,40){\line(1,-1){15}}
\put(70,10){\line(1,1){30}}
\put(70,40){\line(1,-1){30}}
\footnotesize
\put(85,7){\makebox(0,0){$r=0$}}
\put(85,43){\makebox(0,0){$r=0$}}
\put(61,16){\makebox(0,0){${\it I}^-$}}
\put(61,34){\makebox(0,0){${\it I}^+$}}
\put(111,16){\makebox(0,0){${\it I}^-$}}
\put(111,34){\makebox(0,0){${\it I}^+$}}
\put(96,34){\makebox(0,0){$r=r_+$}}
\put(96,16){\makebox(0,0){$r=r_+$}}
\put(75,25){\makebox(0,0){\mbox{${\bf I}$}}}
\put(95,25){\makebox(0,0){\mbox{${\bf I}$}}}
\put(85,31){\makebox(0,0){\mbox{${\bf I}$}\mbox{${\bf I}$}}}
\put(85,19){\makebox(0,0){\mbox{${\bf I}$}\mbox{${\bf I}$}}}
\end{picture}

\noindent Fig. 7

Penrose diagram for the asymptotically flat uncharged \\ black
hole in even dimensions. \\
\end{center}

It should be stressed here that the metric (\ref{f5}) is not an
extremum for the Hilbert action but, rather, for the action

\begin{equation}
I_{l \mbox{$\rightarrow $}\infty} = \frac{1}{4(\mbox{$\cal
D$}-2)!\Omega_{{\cal D}-2}}\int
\epsilon_{a_1,...,a_{{\cal D}}} R^{a_1 a_2} \mbox{\tiny $\wedge$}
\cdots \mbox{\tiny $\wedge$} R^{a_{\mbox{\tiny $\cal D$}-3}
a_{\mbox{\tiny $\cal D$}-2}} \mbox{\tiny $\wedge$}
e^{a_{\mbox{\tiny $\cal D$}-1}} \mbox{\tiny $\wedge$}
e^{a_{\mbox{\tiny $\cal D$}}}
\label{n4}
\end{equation}
in which only the term with the highest non trivial power of the
curvature tensor in the Lovelock action is retained.


\section{Thermodynamics}

\subsection{ Temperature }

The black-hole temperature can be calculated by imposing
regularity at the horizon of the Euclidean continuation of the
manifold \cite{GH}. In imaginary time
$t=-i\tau$, the black-hole metric (\ref{3.1}) takes the form

\begin{equation}
ds^2_{Euc} = N^2(r) g^2(r) d\tau^2 + g^{-2}(r) dr^2 + r^2
d\Omega^2
\label{t1}
\end{equation}
where $g^2(r)$ is given by (\ref{3.6}) and $N(r)=N_{\infty}$. The
Euclidean section is defined for $r>r_+$. Since $g^2(r_+)=0$,
the Euclidean black hole has the topology $\Re^2\times
S^{\mbox{\tiny $\cal D$}-2}$.

The Euclidean coordinate $\tau$ is periodic and one may fix its
range in the interval [0,1].  If this is done, $N_{\infty}$ represents
the Euclidean time period whose inverse is the temperature: $N
\equiv \beta = T^{-1}$. The standard formula

\begin{equation}
T=\frac{1}{4\pi} \left( \frac{d g^2(r)}{dr} \right)_{r=r_+}
\label{t9}
\end{equation}
relates the inverse Euclidean period with the black-hole
parameters (mass and electric charge) so that no conical
singularity appears at the horizon.  In the case $Q=0$ one
obtains,

\begin{equation}
T = \left\{ \begin{array}{lll} \frac{1 + (2n-1)(r_+/l)^2 }{4\pi
(n-1)r_+}, & \;\;\;\;\;\; \mbox{$\cal D$}=2n & \;\;\;(a) \\ r_+/(2\pi l^2)
\;, & \;\;\;\;\;\; \mbox{$\cal D$}=2n-1 & \;\;\;(b)
\end{array} \right.
\label{t10}
\end{equation}
This last relation shows that evaporating black holes would
behave very differently for even and odd \mbox{$\cal D$}. As
$M\rightarrow 0$, $r_+ \rightarrow 0$ in both cases, whereas
$T\rightarrow
\infty$
for $\mbox{$\cal D$}=2n$ and $T\rightarrow 0$ for $\mbox{$\cal
D$}=2n-1$. In the limit $l\rightarrow \infty$, (\ref{t10}a)
reproduces the standard result for
\mbox{$\cal D$}=4, $T =(8\pi m)^{-1}$. For $n\rightarrow \infty$,
(\ref{t10}a) approaches (\ref{t10}b).


\subsection{Euclidean Action and Entropy}

A deeper insight into black hole thermodynamics is gained from
the identification of the Euclidean path integral in the saddle
point approximation around the black-hole solution with the
partition function for a thermodynamic ensemble\cite{GH}. In
this approximation,

\begin{eqnarray}
I_{Euc} &=& \frac{\mbox{Free Energy}}{\mbox{Temperature}}
\nonumber\\ &=& \frac{M}{T} - S +
\sum \frac{\mu_i}{T} Q^i
\label{5.15}
\end{eqnarray}
where the $\mu_i$ are the chemical potentials associated with
the charges $Q^i$.

Consider the reduced action of the coupled system (\ref{q5})

\begin{eqnarray}
I &=& I_G + I_Q \nonumber \\ &=& (t_2-t_1) \int dr \left[ N \left(F'-
\frac{1}{2} r^{\mbox{\tiny $\cal D$}-2} p^2 \right) + \varphi
(r^{\mbox{\tiny $\cal D$}-2}p)' \right] + B.
\label{5.16}
\end{eqnarray}

The Euclidean and Minkowskian actions are related by

\begin{equation}
e^{iI_M}=e^{-I_E},\;\;\;\;\; \tau=it.
\label{5.17}
\end{equation}
For the thermodynamical description of the black hole, one needs
the Euclidean action evaluated on the physical values of the
fields (i.e., those which solve the field equations with
Minkowskian signature). Thus, one obtains

\begin{equation}
I_E = -\int_{r_+}^{\infty} dr \left[ N \left(F'- \frac{1}{2}
r^{\mbox{\tiny $\cal D$}-2} p^2 \right) +
\varphi (r^{\mbox{\tiny $\cal D$}-2}p)' \right] + B_E.
\label{5.18}
\end{equation}
where we have set $\tau_2-\tau_1=1$.

Since the volume piece of the reduced action is a linear
combination of the constraints, the on-shell value of the action
is equal to the boundary term $B_E$. The value of $B_E$, which
is fixed by the boundary conditions gives then the free energy
of the system\cite{Action}

\begin{equation}
B_E = \frac{M}{T} + \sum \frac{\mu_i}{T} Q^i - S.
\label{5.19}
\end{equation}

Let us compute $B_E$. For $r \mbox{$\rightarrow $}\infty$, we
have

\begin{equation}
\begin{array}{rcl}
N &\rightarrow& \mbox{$N_{\infty}$} \equiv \beta \\
\varphi  &\rightarrow& \varphi_{\infty} \\
\delta F &\rightarrow& \delta M \\
\delta p &\rightarrow& r^{-\mbox{\tiny $\cal D$}+2} \delta Q.
\end{array}
\label{5.20}
\end{equation}
At the horizon, we impose the regularity condition of no conical
singularities (\ref{t9})

\begin{equation}
N (r_+) \left( \frac{\mbox{$\partial $} g(r)^2}{\mbox{$\partial
$} r}\right)_{r=r_+} = 4\pi
\label{5.21}
\end{equation}
and the condition

\begin{equation}
\varphi(r_+) = 0.
\label{5.22}
\end{equation}
This last condition is just a matter of convention. One can
assume a non-zero value of $\varphi$ at the horizon and impose
instead $\varphi(\infty)=0$. The difference between these choices
is that the chemical potential associated to the electric charge
is, in the first case, the value of $\varphi$ at infinity, while
in the second case, it is minus the value of $\varphi$ at
the horizon.  Both descriptions are, however, equivalent
\cite{f4}.

Once the boundary conditions have been chosen, it remains to
choose $B_E$ so that it cancels all the boundary terms that
appear by partial integration in the variation of the Euclidean
action.

The variation of the Euclidean action is
\begin{equation}
\delta I_E = -\left[N\delta F + \varphi r^{-\mbox{\tiny $\cal D$}+2}
\delta p \right]_{r_+}^{\infty} + \delta B_E(\infty) + \delta
B_E(r_+) + \mbox{terms vanishing on shell}.
\label{5.23}
\end{equation}

Let us compute first the term coming from the horizon.  As the
value of $\varphi$ is zero there, the only contribution to $B_E$
from $r_+$ is the gravitational term

\begin{equation}
\delta B_E(r_+) = - N(r_+) (\delta F)_{r_+}.
\label{5.24}
\end{equation}

The variation of $F$ at the horizon is given by

\begin{equation}
(\delta F)_{r=r_+} = \left(\frac{\mbox{$\partial $}
F}{\mbox{$\partial $} g^2}\right)_{r=r_+} [\delta g^2
(r)]_{r=r_+}
\label{5.25}
\end{equation}
and, from the definition of the horizon $g^2(r_+)=0$, it follows
that

\begin{equation}
[\delta g^2]_{r_+} + \left(\frac{d g^2}{dr}\right)_{r=r_+}
\delta r_+ = 0,
\label{5.26}
\end{equation}
then, one finds,

\begin{eqnarray}
\delta B_E(r_+) &=& N(r_+) \left(\frac{d g^2}{dr}\right)_{r=r_+}
\left(\frac{\mbox{$\partial $} F}{\mbox{$\partial $}
g^2}\right)_{r=r_+}\delta r_+ \nonumber\\
&=& 4\pi \left(\frac{\mbox{$\partial $} F}{\mbox{$\partial $}
g^2}\right)_{r=r_+} \delta r_+
\label{5.27}
\end{eqnarray}
where we have used the condition (\ref{5.21}).

Let us now compute the term coming from the asympototic region
$B_E(\infty)$.  Using the boundary conditions (\ref{5.20}) and
(\ref{5.23}) one easily finds

\begin{equation}
\delta B_E(\infty) = \beta \delta M + \varphi_{\infty} \delta Q
\label{5.28}
\end{equation}

Thus, as the value of the boundary term $B_E \equiv
B_E(\infty)+B_E(r_+)$ is equal to the free energy, one finds

\begin{equation}
\mbox{Free energy} = \beta M + \varphi_{\infty} Q + 4\pi \int dr_+
\left(\frac{\mbox{$\partial $} F}{\mbox{$\partial $}
g^2}\right)_{r=r_+} - S_0. \label{5.29}
\end{equation}
where $S_0$ is a constant without variation. Comparing
(\ref{5.29}) with (\ref{5.19}) one learns that $M$ is the
internal energy of the system, $\beta$ is the inverse
temperature and $\beta^{-1}\varphi_{\infty}$ is the chemical
potential conjugate to the electric charge. The entropy as a
function of $r_+$ is given by

\begin{equation}
S(r_+)= - 4\pi \int dr_+ \left(\frac{\mbox{$\partial $}
F}{\mbox{$\partial $} g^2(r)}\right)_{r=r_+} + S_0
\label{5.30}
\end{equation}
and the black-hole radius $r_+$ depends on the mass and electric
charge through the equation

\begin{equation}
g^2(r_+)=0 \;\;\; \mbox{$\rightarrow $}\;\;\; r_+=r_+(M,Q)
\label{5.31}
\end{equation}

In even dimensions, the integral (\ref{5.30}) can be computed in
closed form obtaining

\begin{equation}
S(r_+) = \pi l^2 \left[ \left( 1 +
\frac{r_+^2}{l^2}\right)^{n-1}-1\right].
\label{5.32}
\end{equation}
Here, we have fixed $S_0=-\pi l^2$ so that

$$S(r_+=0)=0.$$ This provides the usual result in \mbox{$\cal
D$}=4 $(n=2)$ and also a finite value for the entropy in the
limit $l\rightarrow \infty$, $S(r_+)\rightarrow \pi (n-1)
r^2_+$.

In odd dimensions, the integral

\begin{equation}
S(r_+)= 4\pi (n-1) \int_0^{r_+} dr_+ \left(1 + \frac{r_+^2}
{l^2}\right)^{n-2} + S_0.
\label{5.33}
\end{equation}
cannot be evaluated in closed form for arbitrary $n$.

Choosing $S_0=0$ gives zero entropy for the vacuum solution
($r_+=0$) and one recovers the value $S=4\pi r_+$ for the
special case \mbox{$\cal D$}=3\cite{BTZ}.  Since the energy is
dimensionless for odd
\mbox{$\cal D$}, the length parameter $l$ plays a key role in
the black-hole properties. The limit $l\rightarrow \infty$ is
not well defined due to the dependence on $l$ of $r_+$ (for
$Q=0$).  When $l\rightarrow \infty$, $r_+$ also goes to infinity
and one is left only with the black-hole interior. In even
dimensions on the other hand, $r_+ \rightarrow 2M$ for $l
\rightarrow \infty$.

The entropy in the higher dimensional black holes described here
is no longer proportional to the area of the horizon. It is,
however, an increasing function of $r_+$ and therefore the second
law of thermodynamics holds.  It should be stressed nevertheless,
that for a generic choice of the Lovelock coefficients, the
entropy is not necessarily a monotonically increasing function
of $r_+$ \cite{14}. Our result for the entropy is a particular
case of the expression given in \cite{14}.  The Lagrangian
method used in that reference, however, involves the
substraction of a divergent term at infinity that obscures the
fact that the entropy is associated with the horizon.
\vspace{1cm}

\noindent {\bf Added Note:} After this paper was written we learned that
Theodore Jacobson and Robert Myers had obtained a general
expression for the entropy of dimensionally continued black
holes. Our results agree with their formula for the special case
of spherical symmetry. We are grateful to Drs.  Jacobson and
Myers for their careful critical reading of this paper.  We thank
them for that and, in particular, for their suggestion that the
singularity at $r=r_c$ in the charged case might be a curvature
singularity, which proved to be correct.

\vspace{2cm}

\begin{center}
{\bf {\large Acknowledgements}}
\end{center}

This work was partially supported by grants 0862/91, and
193.0910/93 from FONDECYT (Chile), by a European Communities
research contract, and by institutional support to the Centro de
Estudios Cient\'{\i}ficos de Santiago, provided by SAREC
(Sweden) and a group of chilean private companies (COPEC, CMPC,
ENERSIS). M.B. holds a Fundaci\'on Andes Fellowship, and J.Z.
wishes to express his gratitude to Prof. Abdus Salam, the
International Atomic Energy Agency and UNESCO for hospitality at
the International Centre for Theoretical Physics, Trieste.


\appendix{Lorentz Lie algebra as a subalgebra of anti-de Sitter algebra}

The anti-de Sitter group in \mbox{$\cal D$}\ dimensions is
isomorphic to the orthogonal group in $\mbox{$\cal D$}+1$
dimensions $SO(\mbox{$\cal D$}-1,2)$. Let
$\eta_{AB}$=diag$(-1,1,...,1,-\mbox{$l^{-2}$})$
$(A,B=0,1,...,\mbox{$\cal D$})$ and $J_{AB}$ are the generators
of the group,

\begin{equation}
\left[ J_{AB} ,J_{CD}
\right] = -J_{AC} \eta_{BD} + J_{AD} \eta_{BC} + J_{BC}\eta_{AD}
- J_{BD}\eta_{AC}.
\end{equation}

The generators $J_{AB}$ can be split into rotations in
\mbox{$\cal D$}\
dimensions (generated by $J_{ab}$) and ``inner translations"
(generated by $J_{a\mbox{\tiny $\cal D$}} \equiv J_a$). The
conmutation relations of the split generators defines the
anti-de Sitter algebra

\begin{eqnarray}
\left[ J_{a} ,J_{b} \right] & = & \mbox{$l^{-2}$} J_{ab} \nonumber \\
{}\left[ J_{ab} ,J_{c} \right] & = & J_{a} \eta_{bc} - J_{b}
\eta_{ac} \label{A1}
\\ {} \left[ J_{ab} ,J_{cd} \right] & = & -J_{ac} \eta_{bd} + J_{ad}
\eta_{bc} +  J_{bc}\eta_{ad} - J_{bd}\eta_{ac}. \nonumber
\end{eqnarray}

Let $W^{AB}$ be the connection one-form for the group and $D=d +
J_{AB} W^{AB}$ the exterior covariant derivative. The curvature
$2-$form is defined as usual

\begin{equation}
\tilde{R}^{AB} = d W^{AB} + W^{A}_{\;C}\mbox{\tiny $\wedge$} W^{CB}.
\label{A2}
\end{equation}

The connection $W^{AB}$ can also be decomposed into the
connection under \mbox{$\cal D$}-rotations $w^{ab}$, and inner
translations $e^a$ in the form

\begin{equation}
W^{AB} = \left( \begin{array}{c|c} w^{ab} & - e^a \\ \hline e^b
& 0
\end{array}  \right).
\end{equation}
In this splitting, the curvature two-form reads

\begin{equation}
\tilde{R}^{AB} = \left(  \begin{array}{c|c}
R^{ab} + \mbox{$l^{-2}$} e^a \mbox{\tiny $\wedge$} e^b & - T^a
\\ \hline T^b & 0 \end{array}
\right)
\label{A4}
\end{equation}
where $T^a \equiv de^a + w^a_{\;b} \mbox{\tiny $\wedge$} e^b$
is the torsion two-form.  Note that $e^a$ transforms as a vector
under rotations generated by $J_{ab}$, but as a connection under
those generated by $J_a$.

\appendix{Hamiltonian Formalism for the Lovelock Action}

The Hamiltonian form of the action (\ref{a1}) is discussed in
\cite{5}. In that approach, the second order formalism is
used. The torsion tensor is set to zero and the connection is
solved in terms of the local frame and their derivatives.  Just
as in \mbox{$\cal D$}=4, the canonical coordinates are the
spatial components of the metric $g_{ij}$, and their conjugate
momenta $\pi^{ij}$. The time components $g_{0\mu}$ are Lagrange
multipliers associated with the generators of surface
deformations, ${\cal H}_{\mu} = (\mbox{${\cal H}$},{\cal H}_i)$.
The action (\ref{a1}) takes the form

\begin{equation}
I=\int (\pi^{ij}\dot{g_{ij}}-N{\cal H}-N^i{\cal H}_i)
d^{\mbox{\tiny $\cal D$}-1}x dt + B.
\label{B1}
\end{equation}

The momenta can be explicitly given in terms of the velocities,

\begin{equation}
\pi^i_j = -\frac{1}{4} \sqrt{g} \sum_{p=0}^{n-1} \frac{\alpha_p}{2p!}
\frac{(\mbox{$\cal D$}-2p)!}{2^p}
\sum_{s=0}^{p-1} C_{s(p)} \delta^{[i_{1} \ldots i_{2p-1} i]}
_{\,[j_{1} \ldots j_{2p-1} j]} \tilde{R}^{j_{1} j_{2}}_{ \:\:
i_{1} i_{2}} \cdots \tilde{R}^{j_{2s-1} j_{2s}}_{ \:\: i_{2s-1}
i_{2s}} K^{j_{2s+1}}_{i_{2s+1}} \cdots K^{j_{2p-1}}_{i_{2p-1}}
\label{B2}
\end{equation}
where $C_{s(p)} = \frac{(-4)^{p-s}}{s![2(p-s)-1]!!}$, and

\begin{equation}
\delta^{[i_1\cdots i_l]}_{[j_1\cdots j_l]} = \left|
\begin{array}{rcl} \delta^{i_1}_{j_1} & \cdots &
\delta^{i_1}_{j_l} \\ \vdots \;\; & & \;\; \vdots \\
\delta^{i_l}_{j_1} & \cdots & \delta^{i_l}_{j_l}
\end{array} \right|.
\label{B3}
\end{equation}

The generators of reparametrizations of the surfaces $t=const$,

\begin{equation}
{\cal H}_{i} = -2\pi^l_{\;i /l},
\label{B4}
\end{equation}
do not depend on the action but only on the transformation
laws of $g_{ij}$ and $\pi^{ij}$. Here $/$ denotes covariant
differentiation in the spatial metric, and $\pi^i_{\;j / i} =
\pi^i_{\;j , i} - \Gamma^n_{ji} \pi^i_n $.

The normal generator ${\cal H}$ is

\begin{equation}
{\cal H} = - \sqrt{det(g_{ij})} \: \sum_{p=0}^{n-1}
\frac{(\mbox{$\cal D$}-2p)!}{2^p} \; \alpha_p \, \delta ^{[i_{1}
\ldots i_{2p}]} _{\,[j_{1} \ldots j_{2p}]}\: \tilde{R}^{j_{1}
j_{2}}_{ \:\: i_{1} i_{2}} \tilde{R}^{j_{3} j_{4}}_{ \:\: i_{3}
i_{4}}\cdots \tilde{R}^{j_{2p-1} j_{2p}}_{ \:\: i_{2p-1} i_{2p}},
\label{B5}
\end{equation}
where the $\tilde{R}^{ij}_{ \:\:kl}$ are the spatial
components of the space-time curvature tensor. They depend on
the velocities through the Gauss--Codazzi equations

\begin{equation}
\tilde{R}_{ijkl} = R_{ijkl} + K_{ik} K_{jl} -  K_{il} K_{jk}.
\label{B6}
\end{equation}
where $R_{ijkl}$ are the components of the intrinsic curvature
tensor of the spatial sections, constructed from $g_{ij}$ and
its spatial derivatives, and $K_{ij} = \frac{1}{2N}
(-\dot{g}_{ij} + N_{i/j}+N_{j/i})$.

In the static, spherically symmetric case one can ignore the
distinction between $\tilde{R}_{ijkl}$ and $R_{ijkl}$ (see eq.
(\ref{B6})). Also, the only non vanishing components of the
spatial curvature are

\begin{equation}
\begin{array}{rcl}
R^{m_1 m_2}_{n_1 n_2} &=& \frac{f(r)}{r^2} \delta^{[m_1
m_2]}_{[n_1 n_2]} \\ R^{r m}_{r n} &=&
\frac{f'(r)}{2r}\delta^{m}_{n}
\label{B9}
\end{array}
\end{equation}
where $f(r) \equiv 1-g^{rr}(r)$.

The normal generator ${\cal H}$ can be computed by direct
substitution of (\ref{B9}) into (\ref{B5}) obtaining

\begin{equation}
\mbox{${\cal H}$}= -(\mbox{$\cal D$}-2)! \sqrt{\gamma} g^{-1}
\frac{d}{dr} \left[ r^{\mbox{\tiny $\cal D$}-1}
\sum \alpha_p (\mbox{$\cal D$}-2p) \left(\frac{1-g^2}{r^2}\right)^p \right],
\label{ap8}
\end{equation}

Replacing in (\ref{ap8}) the choices (\ref{1.3}) for the
coefficients $\alpha_p$ the expression (\ref{3.3}) for the
function $F[r,g(r)]$ is obtained.

The tangential generator \mbox{${\cal H}$}$_i$ is identically
zero for the metrics considered here.

In the transition from (\ref{B5}) to (\ref{ap8}) the following
identities between the antisymmetrized Kroneker deltas are
useful. For $m<p$

\begin{eqnarray}
\delta^{[i_1 \dots i_p]}_{[j_1 \dots j_p]}
\delta^{j_1}_{i_1}\delta^{j_2}_{i_2} \cdots \delta^{j_m}_{i_m}
&=& \frac{(r-p+m)!}{(r-p)!}
\delta^{[i_{m+1} \dots i_p]}_{[j_{m+1} \dots j_p]}
\\
\delta^{[i_1 \dots i_{2p}]}_{[j_1 \dots j_{2p}]}
\delta^{[j_1 j_2]}_{[i_1 i_2]} \cdots \delta^{[j_{2m-1}
j_{2m}]}_{[i_{2m-1} i_{2m}]} &=& \frac{2^m(r-2[p-m])!}{(r-2p)!}
\delta^{[i_{2m+1} \dots i_{2p}]}_{[j_{2m+1} \dots j_{2p}]}.
\end{eqnarray}
where $r$ is defined as the range of the indices.

\newpage



\begin{thebibliography}{10}
\bibitem{f1}  Throughout this work we define the integer $n$ by
$\mbox{$\cal D$}=2n$ for even dimensions, and $\mbox{$\cal
D$}=2n-1$ for odd dimensions.
\bibitem{f2}  Here we will assume that the torsion tensor is not
explicitly contained in the action, the extension of Lovelock
theory for this case is worked out in ref.  \cite{2}.
\bibitem{f3} In odd dimensions without cosmological constant,
a cosmological horizon appears when $Q \neq 0$.  However, the
singularity is then naked so we will not consider that geometry
here.
\bibitem{Action} If one considers the unreduced action, this
argument still holds. In fact, the general form of the action of
the coupled system is $I= \int( \pi^A \dot{\Phi}_A
+ \lambda^A G_A) + B$, where $B$ is a boundary term, the
generators $G_A$ vanish on shell, and $\pi^A, \Phi_A$ denote the
gravitational and electromagnetic fields. If one expects to have
a system in thermal equilibrium, the solution must be static
($\dot{\Phi}_A=0$) and hence, the value of the
action is given by the boundary term $B$ alone.
\bibitem{f4}  The one-form $A_{\mu} dx^{\mu} = \varphi d\tau$ is
singular at $r=r_+$ if $\varphi (r_+)\neq 0$. This singularity
is a gauge artifact and is removed by the (singular) gauge
transformation $\varphi \rightarrow \varphi - \varphi (r_+)\tau$
leading to (\ref{5.22}). Conceptually it is preferable to use
(\ref{5.22}) which makes it explicit that the electric charge is
related to the gauge invariance at infinity.  One may also
consider $\varphi(r_+)=0$ as fixing the ``initial coordinate"
$\varphi(r_+)$, and the ``final momentum" $p(\infty)$.  This
amounts to setting $\delta Q=0$ and hence no surface integral
proportional to $Q$ comes in. The free energy thus obtained is
that of the canonical ensemble $F=U-TS$, and not the grand
canonical one, $F=U-TS-\mu Q$.
\end{thebibliography}
\end{document}